\documentclass[11pt, oneside]{article}   	
\usepackage{geometry}                		
\usepackage{graphicx}				
\usepackage{xcolor,cite}	
\usepackage{multirow,multicol,hyperref}
\hypersetup{colorlinks,bookmarksopen,bookmarksnumbered,
	linkcolor=red!50!black,pdfstartview=FitH,urlcolor=red!30!black,citecolor=green!50!black}
\usepackage[framemethod=tikz]{mdframed}
\usepackage{mathrsfs}
\usepackage[most]{tcolorbox} 							
\usepackage{amsmath}	
\usepackage{amsfonts}	
\usepackage{bbm}
\usepackage[toc,page]{appendix}						
\usepackage{arydshln}

\newcommand{\nn}{\nonumber}
\newcommand{\be}{\begin{equation}}
\newcommand{\ee}{\end{equation}}
\newcommand{\bea}{\begin{eqnarray}}
\newcommand{\eea}{\end{eqnarray}}

\renewcommand{\Re}{{\rm Re}\,}
\renewcommand{\Im}{{\rm Im}\,}
\newcommand{\bp}{\bar{M}_{\rm Pl}}
\def\beq{\begin{equation}}
\def\eeq{\end{equation}}
\newcommand{\fig}[1]{~\ref{fig:#1}}
		
\font\ital=cmu9
\makeatletter
\def\hhref#1{\href{http://arxiv.org/abs/#1}{arXiv:#1}}
\usepackage{xstring}
\newcommand{\hhrefq}[1]{\IfSubStr{#1}{:}{\href{http://inspirehep.net/search?ln=en&ln=en&p=#1&of=hb&action_search=Search&sf=&so=d&rm=&rg=25&sc=0}{InSpire:#1}}{\hhref{#1}}}

\def\art{\@ifnextchar[{\eart}{\oart}}
\def\eart[#1]#2#3#4#5#6{{\rm #2}, {\em #3 \bf #4} {\rm (#6) #5} ({\em #1})}
\def\article{\@ifnextchar[{\earticle}{\oarticle}}
\def\oarticle#1#2#3#4#5#6{{\rm #1}, {\ital `#6'}, {\rm #2 #3 (#5) #4}}
\def\earticle[#1]#2#3#4#5#6#7{{\rm #2}, {\ital `#7'}, {\rm #3 #4 (#6) #5}  [\hhrefq{#1}]}
\def\hepart[#1]#2{{\rm #2, \sl#1}}
\def\heparticle[#1]#2#3{#2, {\ital `#3'} [\hhrefq{#1}]}
\newcommand{\doi}[1]{\href{http://dx.doi.org/#1}{[link]}}
\newcommand{\eq}[1]{~{\rm (\ref{eq:#1})}}

\newcommand{\hhrefqq}[1]{\IfBeginWith{#1}{10.}{\href{https://doi.org/#1}{doi:#1}}{\hhrefq{#1}}}
\def\earticle[#1]#2#3#4#5#6#7{{\rm #2}, {\ital `#7'}, {\rm #3 #4 (#6) #5}  [\hhrefqq{#1}]}

\begin{document}

\title{
 {\Large\LARGE\bf\color{red!30!black} Solving the strong CP problem in string-inspired theories with modular invariance 
}}
\date{}

\author{
Ferruccio~Feruglio$^{1}$\thanks{E-mail: \href{mailto:feruglio@pd.infn.it}{feruglio@pd.infn.it}},
 Antonio Marrone$^{2,3}$\thanks{E-mail: \href{mailto:antonio.marrone@ba.infn.it}{antonio.marrone@ba.infn.it}},
Alessandro Strumia$^{4}$\thanks{E-mail:  \href{mailto:alessandro.strumia@unipi.it}{alessandro.strumia@unipi.it}},
Arsenii Titov$^{5,1}$\thanks{E-mail: \href{mailto:arsenii.titov@unipd.it}{arsenii.titov@unipd.it}}
\
\\[2ex]
\centerline{
\begin{minipage}{\linewidth}
\begin{center}
$^1${\small\em INFN, Sezione di Padova, Italia}\\
$^2${\small\em Dipartimento Interateneo di Fisica, Bari, Italia}\\
$^3${\small\em INFN, Sezione di Bari, Italia}\\
$^4${\small\em Dipartimento di Fisica, Universit\`a di Pisa, Italia}\\
$^5${\small\em Dipartimento di Fisica e Astronomia, Universit\`a di Padova, Italia}
\end{center}
\end{minipage}}
\\[8mm]}
\maketitle
\thispagestyle{empty}

\centerline{\large\bf Abstract}
\begin{quote}
\indent\large
We show that solutions to the strong CP problem based
on modular invariance can be extended to incorporate
features that appear in string compactifications:
quarks with mostly positive modular weights and non-trivial gauge kinetic functions.
This requires assuming that singularities and zeroes only appear at special points, such as decompactification limits.
We discuss the impact of these assumptions on string gauge unification.
\end{quote}

\tableofcontents
\newpage

\noindent
\section{Introduction}
Modular invariance provides a new solution to the strong CP problem 
under the assumption that the modular chargers of quarks sum to 0;
that CP is only broken by the vacuum expectation value of a modulus $\tau$;
that supersymmetry is present at such high energy scales, possibly around the Planck scale~\cite{Feruglio:2023uof,2404.08032,2406.01689}.
Including extra quarks in a quantum field theory (QFT) context, the quarks that acquire heavy masses
tend to have positive modular weights, so that integrating them out leaves light Standard Model (SM) quarks with mostly {\em negative} modular charges.

Modular invariance and supersymmetry are motivated by super-string compactifications. 
However, in this context the light quarks that appear
in the effective field theory below the string scale tend to have mostly {\em positive} modular charges.
The difference with QFT arises because the heavy quarks in towers of string modes 
fill modular representations with infinite dimension,
transforming into each other under modular transformations,
rather than having a few components with definite modular weights.

\smallskip

We here generalise the modular solution to the strong CP problem to the effective field theories motivated by strings,
with non-negative modular weights, and where  integrating out the string modes leads to 
non-trivial gauge kinetic functions $f_{1,2,3}(\tau)$ of various moduli fields $\tau$ 
for the three  factors in the SM gauge group $G = {\rm SU}(3)\otimes\,{\rm SU}(2)\otimes\,{\rm U}(1)$.
It was discovered that, in a class of simple string compactifications that respect a full modular ${\rm SL}(2,\mathbb{Z})$ invariance,
the string-theory gauge kinetic functions $f_{1,2,3}(\tau)$ can be univocally derived from QFT consistency arguments,
knowing the modular charges of the particles appearing in the QFT below the string scale,
together with the assumption that singularities only appear in de-compactification limits such as $\tau\to i \infty$~\cite{Kaplunovsky:1995jw}.
While our solution to the strong CP problem can be more general, we focus on this context.\footnote{For further works on the solutions to the
strong CP problem based on spontaneous CP violation, see e.g.~\cite{1307.0710,1412.3805,1506.05433,2105.09122,2106.09108,2406.01260,2407.14585,2407.18161,2408.12146,2505.05142}.
See~\cite{2405.18813} for a discussion of how CP can be spontaneously broken in string compactifications.
}

\medskip

We try briefly outlining the mechanism in this introduction. 
We use standard notations, precisely defined later in section~\ref{sec:th}.
The strong CP problem consists in explaining why the physical combination, invariant under rephasing of fermion fields,
\beq\bar\theta=\theta_3+\arg\det m_q\eeq
is small while the CKM phase is large, despite that they both arise from quark mass matrices $m_q$.
Assuming that Higgs doublets $H_{u,d}$ acquire CP-conserving vacuum expectation values,
the phase of quark masses is related to the more fundamental Yukawa matrices as
$\arg\det m_q=\arg(\det Y_u\det Y_d)$.
Furthermore, in a  supersymmetric theory, 
the theta angles are given by the imaginary part of the gauge kinetic functions, as $\theta_3 =-8\pi^2\Im f_3$.
Here $f_3$ is the gauge kinetic function of the SU(3) gauge group; 
$Y_{u,d}$ are the matrices of the Yukawa couplings in the up and down quark sector;
we collectively denote as $\tau$ the moduli, described as chiral super-fields.
As long as supersymmetry is exact, $\bar\theta$ does not depend on the K\"ahler potential and
can be extracted from the part of the supersymmetric Lagrangian
that has an analytic dependence on the chiral superfields. 
The physical combination $\bar\theta$ generalises into a physical combination of super-fields,
invariant under rephasings of chiral super-fields
 \begin{align}\label{Aditau}
\bar\theta=\arg A(\tau)\qquad\hbox{where}\qquad
A(\tau)=e^{ -8\pi^2 f_3(\tau)}\det Y_u(\tau)\det Y_d(\tau).
\end{align}
The super-function $A(\tau)$ only depends on moduli $\tau$ and not on their conjugates $\bar\tau$.
To solve the strong CP problem, we look for conditions that force $A(\tau)$ to be a real, positive constant,
while allowing enough flexibility in $Y_{u,d}(\tau)$ to reproduce the quark masses and mixings, including the CKM phase. 

This can occur if the function $A(\tau)$ is sufficiently symmetric,
as analytic functions with too much symmetry must be constant.

The simplest possibility is that $A(\tau)$ depends on a single complex variable $\tau$,
embedded in a chiral gauge-invariant super-multiplet. 
$A(\tau)$ is necessarily constant if it is invariant under a
shift symmetry $\tau\to \tau +c$
or under a rescaling $\tau\to c \tau$. 
%
If the theory is also CP-invariant, with CP spontaneously broken to deliver the CKM phase, then this constant
should be real. If, in addition, the constant is positive, we get $\arg A(\tau)=0$.
The shift could be implemented assuming that $\tau$ is the relative phase among two
scalars that spontaneously break a U(1) flavour symmetry, using some extra model-building ingredients~\cite{2406.01689}.

\medskip

Such extra ingredients are automatically incorporated in modular invariant supersymmetric theories,
where modular invariance plays the role of flavour symmetry. 
The function $A$ transforms with modular weight  
\beq k_A = k_{f_3} + \sum_{i=1}^3 (2k_{Q_i}+ k_{U_i} + k_{D_i})\eeq
where the det sums the modular weights of the quarks involved in the Yukawas~\cite{Feruglio:2023uof},
and string threshold corrections to $f_3$ add the extra $k_{f_3}$ contribution.
We here ignored Higgs modular charges and super-gravity effects, discussed in the main text.
In the limit $k_{f_3}=0$, where the QCD kinetic function is constant, the condition $k_A=0$
reproduces the solutions to the strong CP problem studied in~\cite{Feruglio:2023uof}. 
We here cover the more general case, also including gauge kinetic functions carrying
a nontrivial dependence on $\tau$, as in many string theory compactifications.

We will identify the conditions under which modular invariance forces $A(\tau)$ to be constant.
In particular, one should assume that $k_A=0$ and that $A(\tau)$ is holomorphic everywhere
in the fundamental domain of the modular group, the point $\tau=i \infty$ included.
Now $A(\tau)$ is the product of a gauge and a Yukawa contribution, that cannot be both holomorphic.
If one of them is holomorphic and has a nontrivial dependence on $\tau$,
it necessarily vanishes somewhere and the other function must develop a corresponding pole
to guarantee the holomorphicity of $A(\tau)$. 
The assumption of $A(\tau)$ being holomorphic everywhere
implies in general the presence of some singularities in the relevant quantities that, as we will see, is actually 
expected in the context of string compactifications.

The string threshold corrections to gauge kinetic functions $f_{1,2,3}$ have been carefully computed in the literature
because their real part is relevant for string gauge coupling unification.
We now found that modular constraints on the imaginary part of $f_3$ can solve the strong CP problem.
In section~\ref{sec:GUT} we discuss the implications of modular weights that can solve the strong CP problem on gauge coupling unification.


\section{Theories with global supersymmetry}\label{sec:th}
\begin{figure}
\centering
\includegraphics[width=8.cm]{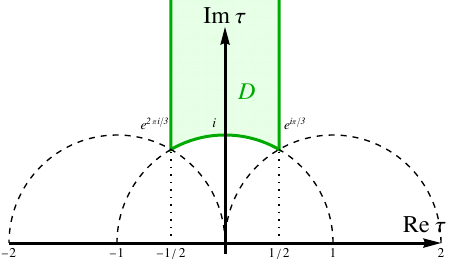}
\caption{\label{FD}\em Fundamental domain $D$ of ${\rm SL}(2,\mathbbm{Z})$. Its closure $\bar D$ is obtained by adding the point $\tau=i\infty$. }
\end{figure}

We consider a ${\cal N}=1$ rigid supersymmetric theory and require invariance under the 
SM gauge group $G = {\rm SU}(3)\otimes\,$SU(2)$\,\otimes\,$U(1) and CP. The field content includes the vector supermultiplets associated to $G$,
three generations of chiral multiplets $M_i=(Q_i,U^c_i,D^c_i,L_i,E^c_i)$ $(i=1,2,3)$ describing quarks and leptons, chiral multiplets $H_u$, $H_d$ describing two electroweak Higgs doublets with opposite hypercharge, 
a pair of gauge invariant supermultiplets $(S,\tau)$, where $\tau$ --- the modulus --- takes values
in the upper half plane
\beq y\equiv -i \tau+i\bar\tau=2\Im\tau >0.\eeq
In a standard superfield notation, the relevant part of the Lagrangian, including the strong and all gauge interactions, reads
\be\label{susylag}
{\cal L}=\int d^2\theta d^2\bar\theta~ K(e^{2V} \Phi,\bar\Phi) + \bigg[\int d^2\theta~ W(\Phi) + \hbox{h.c.}\bigg] +
\bigg[\frac{1}{16}\int d^2\theta~ f_a(\Phi)  W^{a\alpha} W^{a}_{\alpha}+\hbox{h.c.}\bigg],
\ee 
where $\Phi$ collectively denotes all chiral supermultiplets, and $W^{a\alpha}$ 
the gauge supermultiplets, with spinorial index $\alpha$.
The K\"ahler potential $K$ is a real gauge-invariant function. The super-potential $W(\Phi)$ and the gauge kinetic functions $f_a(\Phi)$ $(a=1,2,3)$, one per each group factor, only depend on $\Phi$ and not on their conjugate.
Throughout this work, we adopt a field basis such that the requirement of CP invariance is satisfied when all the 
coupling constants are real. 
CP is spontaneously broken when the VEV of $S$ has a non-vanishing imaginary part and/or the VEV of $\tau$ has a non-vanishing real part.\footnote{To spontaneously break CP, the VEV of $\tau$ should not lie on the boundary of the fundamental domain $D$ of the modular group~\cite{1905.11970}, see fig.~\ref{FD}.}
We also impose that the theory is invariant under the homogeneous modular group $\Gamma={\rm SL}(2,\mathbb{Z})$, whose generic element is
\be
\label{sl2z}
\gamma=
\left(
\begin{array}{cc}
a&b\\c&d
\end{array}
\right)\,,
\ee
where $a$, $b$, $c$ and $d$ are integers obeying $ad-bc=1$.
The modular group acts on the chiral and the vector supermultiplets $(\Phi,V)$ as
\begin{align}
\label{mt}
S\xrightarrow{\gamma}S,\qquad
\tau\xrightarrow{\gamma} \gamma\tau=\frac{a\tau+b}{c\tau+d},\qquad
M_i\xrightarrow{\gamma} (c\tau+d)^{- k_{M_i}} M_i,\qquad
V\xrightarrow{\gamma}V\,.
\end{align}
A possible $\tau$-independent multiplier phase factor in the transformation law of $M_i$ has been omitted.
We assume the action 
\begin{align}
\label{KandW}
K=&~K_S(S)-h^2 \ln y+\sum_{M,i}  \frac{1}{c_{M_i}^2} \frac{\bar M_i  M_i}{y^{k_{M_i}}}\\
W=&~U^c_i  Y^u_{ij}(\tau) Q_j \, H_u  + D^c_i Y^d_{ij}(\tau) Q_j\, H_d  +E^c_i Y^e_{ij}(\tau) L_j\, H_d +
\label{eq:Wgen}
\frac{C^\nu_{ij}(\tau)}{2\Lambda_L} L_i H_u  L_j H_u + \cdots\\
f_a=&~f_a(S,\tau)\,,
\end{align}
where  $h^2, c^2_{M_i}>0$ and $\cdots$ denote additional terms, such as the $\mu$ term and terms depending solely on $(S,\tau)$.
In eq.~(\ref{KandW}), $K_S$ is the part of the K\"ahler potential carrying the dependence on the chiral multiplet $S$.
We do not need to specify it now.
The K\"ahler potential $K$ is invariant up to a K\"ahler transformation, while
invariance of the superpotential requires that
the matrices of Yukawa couplings $Y^{u,d,e}_{ij}(\tau)$ transform as modular functions 
\be
Y^{f}_{ij}(\gamma\tau)=(c\tau+d)^{k_{Y^{f}_{ij}}} Y^{f}_{ij}(\tau),\qquad f = \{u,d,e\}
\ee
with the appropriate weight:
\beq
\label{mw}
k_{Y^{u}_{ij}}=k_{U^c_i}+k_{Q_j}+k_{H_u},\qquad
k_{Y^{d}_{ij}}=k_{D^c_i}+k_{Q_j}+k_{H_d},\qquad
k_{Y^{e}_{ij}}=k_{E^c_i}+k_{L_j}+k_{H_d}\,.
\eeq
The determinants $\det Y^{u,d,e}(\tau)$ are  modular functions with weights $k_{Y^f}=\sum_i k_{Y^{f}_{ii}}$:
\beq
\label{ndet}
\det Y^{f}(\gamma\tau)=(c\tau+d)^{k_{Y^f}} \det Y^{f}(\tau)\,.
\eeq
Under the modular transformations of eq.s~(\ref{mt}), the gauge kinetic functions $f_a(S,\tau)$ 
undergo the change
\begin{align}
\label{anom}
f_a(S,\tau)\to f_a(S,\gamma\tau)-\frac{1}{8\pi^2}\sum_M 2 T_a(M) k_M \ln (c\tau+d)\,,
\end{align}
where the first term comes from the dependence of $f_a(S,\tau)$ on the modulus and the second term 
is due to the change of the path-integral measure, which in general is not left invariant under modular transformations.
Here $T_a(M)$ is the Dynkin index of the $M$ representation under the gauge group of interest. We use the normalization
$T(N) = 1/2$ for the fundamental of SU$(N)$. 
Eq.~(\ref{anom}) signals a potential anomaly of the theory. 
To achieve invariance of the theory and cancel the anomaly, $f_a(S,\tau)$ should satisfy the relation
\be
\label{trf}
f_a(S,\gamma\tau)= f_a(S,\tau)+\frac{1}{8\pi^2}\sum_M 2 T_a(M) k_M \ln (c\tau+d)\,,
\ee
that can be rewritten as
\begin{align}
\label{ef}
e^{ -8\pi^2 f_a(S,\gamma\tau)}=(c\tau+d)^{ k_{f_a}}~e^{ -8\pi^2 f_a(S,\tau)},\qquad
k_{f_a}=-\sum_M 2 T_a(M) k_M\,,
\end{align}
\noindent
showing that $e^{ -8\pi^2 f_a(S,\tau)}$ is a modular function with weight $k_{f_a}$.
For the matter content of the MSSM and its gauge group, these weights are:
\begin{align}
\label{nf}
k_{f_3}=&-\sum_i (k_{U^c_i}+k_{D^c_i}+2 k_{Q_i})&& {\rm SU}(3)\nn\\
k_{f_2}=&-\sum_i (3k_{Q_i}+k_{L_i})-(k_{H_u}+k_{H_d})&& {\rm SU}(2)\\
k_{f_1}=&-\frac13\sum_i (k_{Q_i}+8k_{U^c_i}+2k_{D^c_i}+3k_{L_i}+6k_{E^c_i})-(k_{H_u}+k_{H_d})&& {\rm U}(1)\nn
\end{align}
Eq.s~(\ref{ef},\ref{nf}) are the conditions for the cancellation of mixed modular-gauge anomalies.
If we consider a $\tau$-independent $f_a$ as a special case, the only way to satisfy eq.~(\ref{ef}) is to have vanishing
$k_{f_a}$, a scenario studied in~\cite{Feruglio:2023uof} in relation to the strong CP problem.

\smallskip

From eq.s~(\ref{mw}), (\ref{ndet}), (\ref{ef}) and (\ref{nf}) we see that the MSSM field content allows
to define a function of $\tau$
\begin{align}
A(S,\tau)=e^{ -8\pi^2 f_3(S,\tau)}\det Y^u(\tau)\det Y^d(\tau)\,,
\end{align}
which is physical (it does not depend on rephasing reparameterizations of quark fields)
and transforms as a modular function:
\be
A(S,\gamma\tau)=(c\tau+d)^{k_A} A(S,\tau)\qquad\hbox{with weight}\qquad
k_A=3(k_{H_u}+k_{H_d})\,.
\ee
We assume that the function $A(S,\tau)$ is defined everywhere in the upper half complex plane $\Im\tau>0$ and, 
being the theory modular invariant, we can restrict $\tau$ to the closure $\bar D$ of the fundamental domain $D$ of $\Gamma$,
shown in fig.~\ref{FD}.

To solve the strong CP problem, we are led to assume that:
\begin{itemize}
\item[1.]
The sum of the weights in the Higgs sector vanishes, such that $A$ is modular invariant:
\be\label{sumh}
k_{H_u}+k_{H_d}=0.
\ee
This can more simply happen if $H_u$ and $H_d$ are modular invariant.

\item[2.]
The function $A(S,\tau)$ has no singularities in the closure $\bar D$ of the fundamental domain of $\Gamma$, which includes the cusp $\tau=i\infty$. It follows that $A(S,\tau)$ is a modular form with vanishing weight, which means it does not depend on $\tau$.
\item[3.]
The chiral super-multiplet $S$ has a real VEV and does not break CP. This implies that $A(S,\tau)$ is a real constant
and we further assume it is positive.
We end up with
\be
\bar\theta=\arg A(S,\tau)=0.
\ee
\end{itemize}
The parameter $\bar\theta$ vanishes in the ultraviolet, independently from
the particular vacuum selected by the modulus $\tau$. If the VEV of $\tau$ spontaneously break CP,
a nontrivial CKM phase is generated while keeping $\bar\theta=0$.

\subsection{Zeros and singularities in the moduli space}
It is important to note that the assumption 2 is rather restrictive and carries strong implications. It suffices to midly relax it and our result is lost. For example, a modular invariant function $A(S,\tau)$, analytic everywhere, except at the point $\tau=i\infty$, is not necessarily a constant in the variable $\tau$.
It can be a polynomial of the Klein invariant\footnote{The Eisenstein forms  $E_{k}(\tau)$ are later defined in eq.\eq{Ek}  
and have  $q$ expansion
$E_4(\tau)=1+240 q+\cdots$ and $E_6(\tau)=1-504 q+\cdots$.}   
\be
j(\tau)=\frac{1728 E_4^3(\tau)}{E_4^3(\tau)-E_6^2(\tau)}=\frac{1}{q}+744+\cdots\qquad \hbox{where}\qquad q=e^{ 2\pi i \tau},
\ee
which has a pole at $\tau=i\infty$ ($q=0$) as only singularity. We can easily build examples where the effective field theory
is not well-behaved at some point of the domain $D$ describing the inequivalent vacua of the theory. 
Consider neutrino masses arising from the seesaw mechanism after integrating
out a set of right-handed neutrinos whose masses depend on some VEVs. If these masses vanish when the VEVs
pick a particular value, at this point of the moduli space neutrino masses develop a singularity and the low-energy theory is no longer
valid. In string theory compactifications, these singularities are common. At specific points of the moduli space
towers of states can become massless and singularities show up in the corresponding low-energy theory. 
According to the distance conjecture~\cite{Ooguri:2006in}, this behaviour is unavoidable: as one moves a super-Planckian distance in the moduli space of scalar fields, an infinite tower of states becomes exponentially light
and the effective field theory description breaks down. 
The distance in field space induced by the  kinetic terms of moduli is the Poincar\'e metric
\be
ds^2=\frac{d (\Re\tau)^2+d (\Im\tau)^2}{(\Im\tau)^2}\,,
\ee
showing that the cusp $\tau=i\infty$ is at infinite distance from any point of the fundamental domain $D$.
The distance conjecture has been verified in different specific string contexts 
and, in particular, those giving rise to modular invariant theories~\cite{Gonzalo:2018guu}.
Thus, assuming $A(\tau)$ holomorphic in the whole region $\bar D$ is in apparent conflict with expectations from string theory compactifications and other effective field theories. 

\smallskip

To see how assumption 2 can be realized, 
while allowing singularities to appear in the low-energy theory, we investigate separately the behaviour of 
$\det Y^u(\tau)\det Y^d(\tau)$ and $e^{ -8\pi^2 f_3(S,\tau)}$. 

Concerning the gauge factor in $A$, we assume
\beq \label{eq:f3string}
f_3(S,\tau) = \kappa_3 S - \frac{k_{f_3}}{8\pi^2}\ln\eta^2(\tau).\eeq
This satisfies the modular invariance constraints discussed in the previous section,
since the Dedekind $\eta(\tau)$ function has modular weight $1/2$.\footnote{The transformation of $\eta$ is modular
up to multiplier phases $(-1)^{1/24}$ that would need an extra technical discussion.
This can be bypassed because we will later only need $\eta^{24}$.}
Furthermore, this is motivated by a class of orbifold heterotic string compactifications with full modular invariance.
In this context $\kappa_3$ is a positive integer known as Kac-Moody level,
and the second term is the perturbative string threshold correction, arising at one loop only.
Such correction is fully determined by demanding that
the effective field theory is modular invariant and that
the only singularity appears at $\tau=i\infty$, which is a de-compatification limit.
This selects the $\eta$ function, reproducing the results of explicit string computations~\cite{Kaplunovsky:1995jw}.\footnote{The $\eta$ function can be understood as arising by summing the usual QFT $\ln M$
over string modes with mass $M=|m+n\tau|$~\cite{Ferrara:1991uz}
\beq \label{eq:etamn}\sum_{(m,n)\neq(0,0)} \ln |m+ n\tau|^2 \simeq \ln|\eta(\tau)|^4.\eeq}
Strings lead to multiple modular symmetries, and each one gives a factor as in eq.\eq{f3string},
such that their singularities in the dual de-compactification limits just reproduce the expected behaviour
of a gauge coupling in 6 dimensions.
We can focus on a single modulus $\tau$.

\medskip

Concerning the Yukawa factor in $A$,
we tentatively assume that $\det Y^u(\tau)\det Y^d(\tau)$
has a nontrivial $\tau$ dependence, and remains well-behaved and holomorphic in the whole region $\bar D$. 
This implies that $\det Y^u(\tau)\det Y^d(\tau)$ is a modular form of positive weight 
\beq k_{Y}=\sum_i (k_{Y^u_{ii}}+k_{Y^d_{ii}}) = \sum_{i=1}^3 ({2}k_{Q_i}+k_{U^c_i}+k_{D^c_i}) +3 (k_{H_u}+k_{H_d}) >0.\eeq
Such a form necessarily has zeros in $\bar D$.
The zeros of modular forms have orders that must satisfy the valence formula 
\begin{align}\label{eq:valence}
m(i\infty)+ \frac{m(i)}{2}  + \frac{m(e^{i 2\pi/3})}{3} + \sum_{\tau \in {\bar D}'} m(\tau) = \frac{k_{Y}}{12}\,,
\end{align}
where $m(\tau)$ are the orders of the zeros at $\tau$,
and the sum runs over points $\tau$ away from $i$, $i \infty$, $e^{i 2\pi/3}$.
Given that $e^{-8\pi^2 f_3}$ has a singularity in the de-compactification limit  $\tau=i\infty$
(as motivated by string compactifications),
to obtain a constant $A$ we need the following assumption, perhaps 
realised in some string compactifications:
{\em quark masses never vanish except in the de-compactification limit  $\tau=i\infty$,
where $\det Y^u(\tau)\det Y^d(\tau)$ has a zero} with some order $m(i\infty) \equiv m$. 
Then, the valence formula eq.\eq{valence}
implies that this can only be realized if $\det Y^u(\tau)\det Y^d(\tau)$
is a modular form of weight $k_{Y}=12m$, easily identified as the $m$-th power of the modular discriminant
cusp form $\Delta(\tau)$ with weight 12, defined as
\be
\Delta(\tau)=\frac{E_4^3(\tau)-E_6^2(\tau)}{1728}=q\prod_{n=1}^{\infty}(1-q^n)^{24}=\eta(\tau)^{24}.
\ee
The modular transformation of $\Delta(\tau)$ involves no multiplier phases, unlike the transformation of $\eta(\tau)$.
With these assumptions $A$ does not depend on $\tau$, as required.
Its gauge factor can be rewritten as
\begin{align}
\label{ansatz}
e^{ -8\pi^2 f_3(S,\tau)}=e^{ -8\pi^2 \kappa_3 S }\Delta(\tau)^{-m}\,,
\end{align}
showing it is a modular function of negative weight $k_{f_3}=-12m$, restricted to be a multiple of 12 by our
assumptions about the Yukawas.
Modular forms with negative weight do not exist, and indeed $e^{ -8\pi^2 f_3(S,\tau)}$ develops a singularity 
in the decompactification limit
\be
f_3(\tau)\xrightarrow{\Im\tau\to\infty}  \kappa_3 S -\frac{m}{4\pi}\Im\tau\,,
\ee
where $\det Y^u(\tau)\det Y^d(\tau)$ vanishes. 
The singularity reproduces the behaviour of a gauge coupling in extra dimensions, and 
its negative sign (implying strong coupling when $f_3\to 0$) does not invalidate the argument.
We kept the dilaton $S$, that in string-theory controls large-volume limit at tree level, 
allowing to define a controlled limit that avoids strong coupling: large $\Im\tau$ and $\Re S$, small $\Im\tau/\Re S$~\cite{Kaplunovsky:1995jw}.

\subsection{Yukawa couplings: an illustrative model}
\label{imod}
To make the previous analysis more concrete, 
we discuss an explicit example of modular-invariant Yukawa matrices that can reproduce all observations,
such that the determinant of the quark Yukawa couplings has a single zero at $\tau=i\infty$
 to solve the strong CP problem.
 We here focus on the simplest possibility, altought probably not string-motivated:\footnote{Semi-realistic string compactifications employ modular sub-groups and twisted states, 
leading to Yukawa entries $Y^q_{ij}$ with multipliers (reproduced by $\vartheta$ modular functions)
that possibly cancel in $\det Y$.} 
Yukawas as modular forms under the full SL$(2,\mathbb{Z})$ group.
To obtain a nontrivial CKM phase, we have to make sure that some entry of $Y^{u,d}(\tau)$ is a modular form of weight 12, a linear combination of $E_4^3(\tau)$ and $E_6^2(\tau)$ possessing a phase that cannot be eliminated via field redefinitions. 
We satisfy the condition of eq.~(\ref{sumh})
by choosing $k_{H_{u}}=k_{{H_d}}=0$.
The simplest possible choice of positive weights for the three quark generations then is
\be
\label{eq:Qweights}
k_{U^c_i}=k_{D^c_i}=k_{Q_i}=(2,4,6)\qquad\hbox{corresponding to }\qquad m=4.
\ee
From eq.s~(\ref{mw}) and (\ref{ndet}), we see that this choice delivers 
determinants $\det Y^{u,d}(\tau)$ that are modular forms of weight 24, but not necessarily proportional to
$\Delta(\tau)^2$. Indeed the most general weight-24 modular form is a linear combination of $E_4^6$, $E_4^3 E_6^2$ and
$E_6^4$, whose generic coefficients do not match those required by $\Delta^2\propto(E_4^3-E_6^2)^2$.
On the other hand, each matrix $Y^{u,d}(\tau)$ comes with nine free parameter, that can be adjusted to get
$\det Y^{u,d}(\tau)\propto \Delta^2$. Among the infinitely many possible choices, we find particularly intriguing the
following:
\begin{align}
\label{deus}
Y^{u,d}(\tau)\propto
\left(
\begin{array}{ccc}
E_4(\tau)&E_6(\tau)&E_8(\tau)\\
E_6(\tau)&E_8(\tau)&E_{10}(\tau)\\
E_8(\tau)&E_{10}(\tau)& E_{12}(\tau)
\end{array}
\right),
\end{align}
where 
\beq \label{eq:Ek}
E_{k}(\tau) = \frac{1}{2\zeta(k)} \sum_{(m,n)\neq(0,0)}\frac{1}{(m + n\tau)^{k}}\eeq
are the Eisenstein series, modular forms of weight $k$. 
Notice that the coefficients multiplying $E_{k}$ in $Y^{u,d}$ are all equal to $1$.\footnote{This could possibly arise
if eq.\eq{Ek} represents a common summation over string modes analogous to eq.\eq{etamn}.}
All Eisenstein forms $E_k$ are polynomial in $E_4$ and $E_6$. 
In particular, $E_8=E_4^2$, $E_{10}=E_4 E_6$ and
$E_{12}=(441 E_4^3 +250 E_6^2)/691$. 
The determinant is $\det Y^{u,d}(\tau)\propto(E_4^3-E_6^2)^2$, as desired,
thanks to the fact that the two coefficients in $E_{12}$ sum to 1, and thanks to the 3 generations.
We assume that the K\"ahler potential of quarks and leptons
\be
K=\sum_{i=1}^3 \left[ 
\frac{|Q_i|^2}{c^{2}_{Q_i}\,y^{k_{Q_i}}}+  
\frac{|U^c_i|^2}{c^{2}_{U^{c}_i}\,y^{k_{U^c_i}}}+  
\frac{|D^c_i|^2}{c^{2}_{D^{c}_i}\,y^{k_{D^c_i}}}+
\frac{|L_i|^2}{c^{2}_{L_i}\,y^{k_{L_i}}}+  
\frac{|E^c_i|^2}{c^{2}_{E^{c}_i}\,y^{k_{E^c_i}}}\right]+\cdots
\ee
contains non-universal positive coefficients $c$,  not restricted by modular invariance.
Moving to the basis where the kinetic terms of quarks are canonical, 
the Yukawa couplings become
\begin{align}
\label{deuscan}
Y^q_{\rm can}=
\left(
\begin{array}{ccc}
c_{Q^{c}_1}  c_{Q_1} y^2 E_4&c_{Q^{c}_1} c_{Q_2}y^3 E_6&c_{Q^{c}_1} c_{Q_3}y^4 E_8\\
c_{Q^{c}_2} c_{Q_1}y^3 E_6&c_{Q^{c}_2} c_{Q_2}y^ 4E_8&c_{Q^{c}_2}c_{Q_3}y^5 E_{10}\\
c_{Q^{c}_3} c_{Q_1}y^4 E_8&c_{Q^{c}_3} c_{Q_2}y^5 E_{10}&c_{Q^{c}_3} c_{Q_3}y^6 E_{12}
\end{array}
\right)
\end{align}
where $Q^c = U^c$ for $q=u$ and $Q^c=D^c$ for $q=d$.
This explicitly shows that K\"ahler corrections leave $\det Y^{u,d}$ proportional to $\Delta^2$.

\subsection{Fitting quarks}
Factorising $c_{U^c_3}c_{Q_3}$, the up Yukawa matrix becomes
\begin{align}
 Y^u_\mathrm{can} &= c_{U^c_3} c_{Q_3} \begin{pmatrix}
 u_{13}\,q_{13}\,y^2 E_4 & u_{13}\,q_{23}\,y^3\,E_6 & u_{13}\,y^4 E_4^2 \\
 u_{23}\,q_{13}\,y^3 E_6 & u_{23}\,q_{23}\,y^4 E_4^2 & u_{23}\,y^5 E_4 E_6 \\
 q_{13}\,y^4 E_4^2 & q_{23}\,y^5 E_4 E_6 & y^6 (441 E_4^3 +250 E_6^2)/691
 \end{pmatrix}\,, 
 \label{eq:DeusYu1} 
\end{align}
where $q_{i3} \equiv c_{Q_i}/c_{Q_3}$, 
$u_{i3} \equiv c_{U^c_i}/c_{U^c_3}$ with $i=1,2$.
We similarly factorise $c_{D^c_3} c_{Q_3}$ from $Y^d_{\rm can}$ and define $d_{i3} \equiv c_{D^c_i}/c_{D^c_3}$.
Thus, the 4 quark mass ratios, the 3 mixing angles and the CKM phase  (8 observables)
depend on 8 real parameters --- 6 real (and positive) ratios $q_{13}, q_{23}, u_{13}, u_{23}, d_{13}, d_{23}$ and the complex $\tau$. 
We vary these parameters to optimise the agreement with the 
8 dimensionless quark observables renormalised around $2\,10^{16}$~GeV, see table~\ref{tab:data}.
\begin{table}[t]
\renewcommand{\arraystretch}{1.5}
\centering
\begin{tabular}[t]{|c|c|}
\hline
Observable & Central value  $\pm 1\sigma$  \\
\hline
$m_u/m_c$ & $(1.93 \pm 0.60) \, 10^{-3}$ \\
$m_c/m_t$ & $(2.82 \pm 0.12) \, 10^{-3}$ \\
$m_d/m_s$ & $(5.05 \pm 0.62) \, 10^{-2}$ \\
$m_s/m_b$ & $(1.82 \pm 0.10) \, 10^{-2}$ \\
\hline
$m_t$/GeV & $87.5 \pm 2.1$ \\
$m_b$/GeV & $0.97 \pm 0.01$ \\
\hline
\end{tabular}
\hspace{5ex}
\begin{tabular}[t]{|c|c|}
\hline
Observable & Central value  $\pm 1\sigma$  \\
\hline
$\sin^2\theta^q_{12}$ & $(5.08 \pm 0.03) \, 10^{-2}$ \\
$\sin^2\theta^q_{13}$ & $(1.22 \pm 0.09) \,10^{-5}$  \\
$\sin^2\theta^q_{23}$ & $(1.61 \pm 0.05) \,10^{-3}$ \\
$\delta_\mathrm{CKM}/\pi$ & $0.385\pm0.017$ \\
\hline
\end{tabular}
\caption{\em\label{tab:data} Values of quark masses and mixings renormalized around $2~10^{16}\,{\rm GeV}$, 
assuming the supersymmetry breaking scale $M_\mathrm{SUSY} = 10\,{\rm TeV}$ and $\tan\beta = 10$~\cite{Antusch:2013jca}.}
\end{table}
%
We employ the results of~\cite{Antusch:2013jca} 
assuming the supersymmetry breaking scale $M_\mathrm{SUSY} = 10$~TeV, 
$\tan\beta = 10$ and negligible supersymmetric threshold corrections.
In addition, we have two overall scales 
$c_{U^c_3}c_{Q_3}$ and $c_{D^c_3} c_{Q_3}$, that can be fixed by reproducing 
the values of $m_t$ and $m_b$ reported in table~\ref{tab:data}.

We minimise
\begin{equation}
 \chi^2_q(\tau,{\wp}_q) = \sum_{i=1}^{8} \left(\frac{\mathcal{O}^q_i(\tau,{\wp}_q) - \mu_i}{\sigma_i}\right)^2\,,
\end{equation}
as function of the  model parameters $\{\Re\tau, \Im\tau\}$ and ${\wp}_q = \{q_{13}, q_{23}, u_{13}, u_{23}, d_{13}, d_{23}\}$,
where $\mu_i$ and $\sigma_i$ are the central values and $1\sigma$ uncertainties 
of the 8 dimensionless quark observables $\mathcal{O}^q_i$.
We find the following best-fit point in the parameter space:
\begin{equation}
 \tau = -0.286 + 1.096 i\,, 
 \label{eq:tauBFquarks}
\end{equation}
\begin{equation}
 %
 q_{13} = 0.037\,, ~ q_{23} = 0.075\,, ~
 u_{13} = 0.035\,, ~ u_{23} = 19.98\,, ~ 
 d_{13} = 3.44\,, ~ d_{23} = 0.203\,.
 %
\end{equation}
For these values of the parameters, $\chi^2_q \approx 0$.
Assuming $\tan\beta = 10$ and 
using $v_u = v \sin\beta$ and $v_d = v \cos\beta$, 
with $v = 174$~GeV, for the overall scales we find
\begin{equation}
 c_{U^c_3} c_{Q_3} = 4.15~10^{-4} \qquad \text{and} \qquad c_{D^c_3} c_{Q_3} = 4.76~10^{-4}\,.
\end{equation}
We see that this simple model can successfully accommodate all quark observables. 
The number of model parameters is equal to the number of observables, 
which is an improvement with respect to the modular-invariant quark models 
able to solve the strong CP problem that were previously constructed 
in the literature~\cite{Feruglio:2023uof,2404.08032,2406.01689}.


\subsection{Fitting leptons}
We further extend the model to the lepton sector, 
assuming the same weight pattern as for quarks, namely,
\begin{equation}
 k_{E^c_i} = k_{L_i} = (2,4,6)\,,
\end{equation}
and the Weinberg operator for neutrino masses. 
The Kähler potential and the super-potential are given in eq.\eq{Wgen},
with $\Lambda_L$ denoting the scale of lepton number breaking.
The matrices $Y^e$ and $C^\nu$ have the same form of eq.~\eqref{deus}.
Assuming that $Y^e_{33} \propto C^\nu_{33} \propto E_{12}$ allows to fit leptonic data, but with a value of
$\tau$ incompatible with the value favoured by quarks.
We allow for an extra free parameter in these entries, i.e., $Y^e_{33} = c^e_{33} E_4^3 + (1-c^e_{33}) E_6^2$, and similarly for $C^\nu_{33}$ with the parameter $c^\nu_{33}$. 
With this assumption, $\det Y^e$ and $\det C^\nu$ remain proportional to  $(E_4^3-E_6^2)^2$, allowing to avoid introducing multipliers.
The form of the neutrino mass matrix can be obtained assuming right-handed neutrinos with a mass matrix of the same form, and integrating them out.
For canonically normalised lepton fields, we have
\begin{align}
 Y^e_\mathrm{can} &= c_{E^c_3} c_{L_3} \begin{pmatrix}
 e_{13}\,l_{13}\,y^2 E_4 & e_{13}\,l_{23}\,y^3 E_6 & e_{13}\,y^4 E_4^2 \\
 e_{23}\,l_{13}\,y^3 E_6 & e_{23}\,l_{23}\,y^4 E_4^2 & e_{23}\,y^5 E_4 E_6 \\
 l_{13}\,y^4 E_4^2 & l_{23}\,y^5 E_4 E_6 & y^6 \left[c^e_{33} E_4^3 + \left(1-c^e_{33}\right) E_6^2\right]
 \end{pmatrix}\,, \\[0.4cm]
 C^\nu_\mathrm{can} &= c_{L_3}^2 \begin{pmatrix}
 l_{13}^2\,y^2 E_4 & l_{13}\,l_{23}\,y^3 E_6 & l_{13}\,y^4 E_4^2 \\
 l_{13}\,l_{23}\,y^3 E_6 & l_{23}^2\,y^4 E_4^2 & l_{23}\,y^5 E_4 E_6 \\
 l_{13}\,y^4 E_4^2 & l_{23}\,y^5 E_4 E_6 & y^6 \left[c^\nu_{33} E_4^3 + \left(1-c^\nu_{33}\right) E_6^2\right]
 \end{pmatrix}\,,
\end{align}
where $l_{i3} \equiv c_{L_i}/c_{L_3}$,  
$e_{i3} \equiv c_{E^c_i}/c_{E^c_3}$ with $i=1,2$.
Thus, the 2 charged lepton mass ratios, 
1 neutrino mass ratio and 3 lepton mixing angles (6 dimensionless observables) 
depend on 6 real constants $\ell_{13}, \ell_{23}, e_{13}, e_{23}, c^e_{33}, c^\nu_{33}$ 
and the complex $\tau$. 
We define the neutrino mass ratio as $\delta m^2/|\Delta m^2|$, 
with $\delta m^2 \equiv m_2^2 - m_1^2$ and $\Delta m^2 = m_3^2 - (m_1^2+m_2^2)/2$. 
For charged lepton masses, we use the values from~\cite{Antusch:2013jca}, 
whereas for the neutrino mass-mixing parameters, we employ the results of a recent global analysis of neutrino oscillation data performed in~\cite{2503.07752}.
For completeness, we provide them in table~\ref{tab:dataleptons}.
\begin{table}[t]
\renewcommand{\arraystretch}{1.5}
\centering
\begin{tabular}[t]{|c|c|}
\hline
Observable & Central value  $\pm 1\sigma$  \\
\hline
$m_e/m_\mu$ & $(4.74 \pm 0.04) \, 10^{-3}$ \\
$m_\mu/m_\tau$ & $(5.88 \pm 0.05) \, 10^{-2}$ \\
\multirow{2}{*}{$\delta m^2/|\Delta m^2|$} & $(2.95 \pm 0.06) \, 10^{-2}$ \\
 & $(2.99 \pm 0.07) \, 10^{-2}$ \\
\hline
$m_\tau$/GeV & $1.293 \pm 0.007$ \\
$\delta m^2$/eV$^2$ & $\left(7.37\pm0.17\right) \, 10^{-5}$ \\
\multirow{2}{*}{$|\Delta m^2|$/eV$^2$} & $(2.495 \pm 0.020) \, 10^{-3}$ \\
& $(2.465 \pm 0.020) \, 10^{-3}$ \\
\hline
\end{tabular}
\hspace{5ex}
\begin{tabular}[t]{|c|c|}
\hline
Observable & Central value  $\pm 1\sigma$  \\
\hline
$\sin^2\theta^\ell_{12}$ & $(3.03 \pm 0.14) \, 10^{-1}$ \\
$\sin^2\theta^\ell_{13}$ & $(2.23 \pm 0.05) \,10^{-2}$  \\
\multirow{2}{*}{$\sin^2\theta^\ell_{23}$} & $(4.73 \pm 0.24) \,10^{-1}$ \\
& $(5.45 \pm 0.23) \,10^{-1}$ \\
\hline
\multirow{2}{*}{$\delta_\mathrm{PMNS}/\pi$} & $1.20 \pm 0.22$ \\
& $1.48 \pm 0.12$ \\
\hline
\end{tabular}
\caption{\em\label{tab:dataleptons} Values of charged lepton masses~\cite{Antusch:2013jca} and 
neutrino mass-mixing parameters~\cite{2503.07752}. 
When two values are present for the same observable, 
the value in the upper (lower) line is for normal (inverted) neutrino mass ordering. 
For neutrino parameters, we use symmetric $1\sigma$ errors defined as 1/6 of the corresponding $3\sigma$ ranges derived in~\cite{2503.07752}.}
\end{table}
%

\begin{figure}[t]
\centering
\includegraphics[width=8.cm]{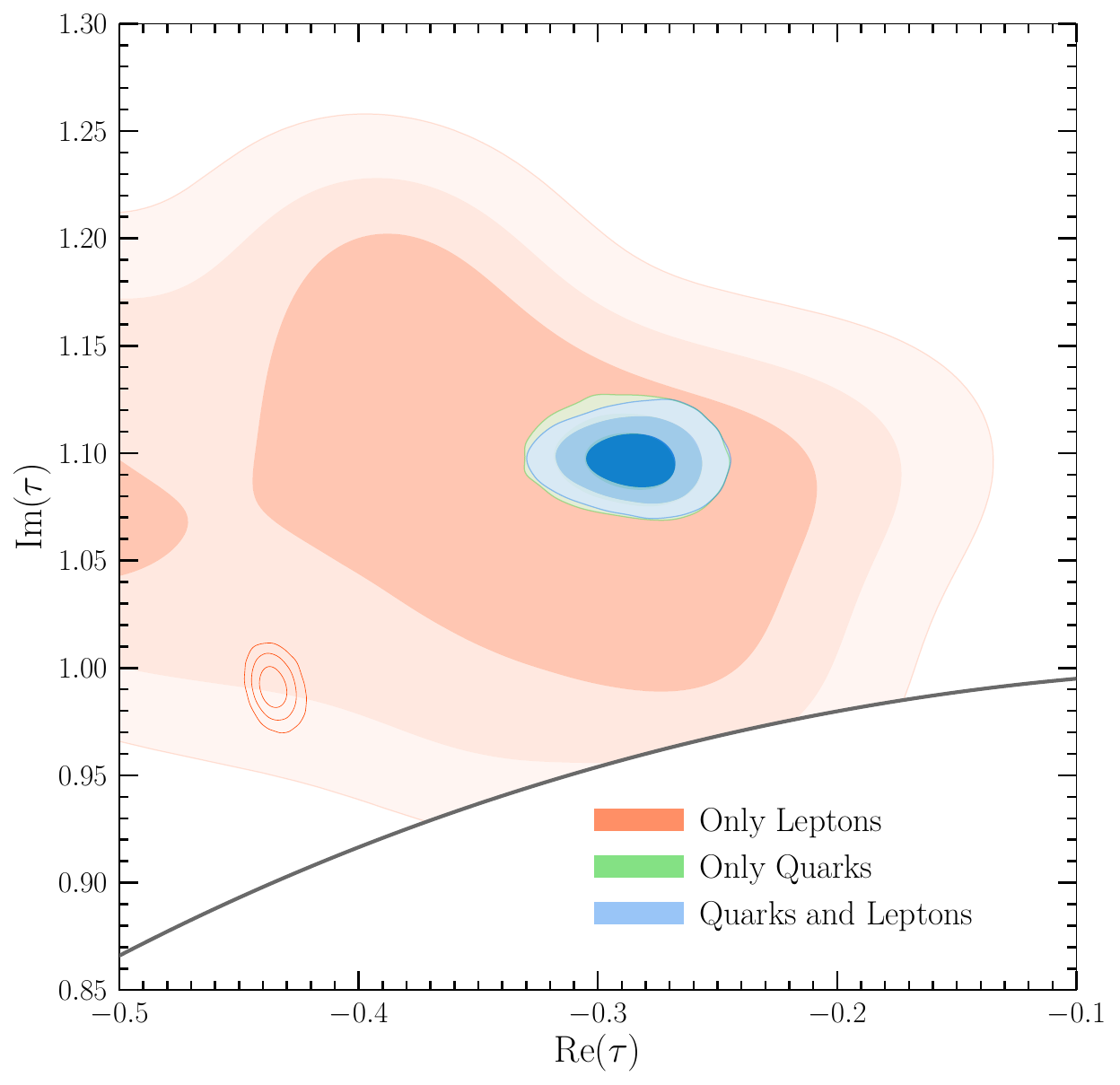}
\caption{\label{fig:tauregions}\em The $1\sigma$, $2\sigma$ and $3\sigma$ regions in the $\tau$ plane obtained by minimising $\chi^2_q$ for quarks only, $\chi^2_\ell$ for lepton only, 
as well as their sum $\chi^2_\mathrm{tot} = \chi^2_q + \chi^2_\ell$. 
{The solid orange lines denote the $1\sigma$, $2\sigma$ and $3\sigma$ contours obtained by minimising $\chi^2_\ell$ for the lepton model with $Y^e_{33} = C^\nu_{33} = E_{12}$.}}
\end{figure}

Fixing $\tau$ to the best-fit value in eq.~\eqref{eq:tauBFquarks} obtained from a fit to the quark data, 
we minimise
\begin{equation}
 \chi^2_\ell(\tau,{\wp}_\ell) = \sum_{i=1}^{6} \left(\frac{\mathcal{O}^\ell_i(\tau,{\wp}_\ell) - \mu_i}{\sigma_i}\right)^2\,
\end{equation}
with respect to the parameters ${\wp}_\ell = \{l_{13}, l_{23}, e_{13}, e_{23}, c^e_{33}, c^\nu_{33}\}$.
Here, $\mu_i$ and $\sigma_i$ are the central values and $1\sigma$ uncertainties 
of the 6 dimensionless lepton observables $\mathcal{O}^\ell_i$. 
Note that we do not include $\delta_\mathrm{PMNS}$ in this set of observables
because of the associated large experimental uncertainty, 
but rather regard the obtained value as a prediction.
We find $\chi^2_\ell \approx 0$ for the following values of ${\wp}_\ell$:
\begin{equation}
 l_{13} = 2.51\,, ~ l_{23} = 1.94\,, ~
 e_{13} = 2.21\,, ~ e_{23} = 0.0079\,, ~ 
 c^e_{33} = 5.61\,, ~ c^\nu_{33} = 0.076\,.
\end{equation}
Assuming $\tan\beta = 10$, we fix the overall scales $c_{E^c_3} c_{L_3}$ and $c_{L_3}^2/2 \Lambda_L$ by reproducing
$m_\tau$ and the neutrino mass squared differences $\delta m^2$ and $\Delta m^2$:
\begin{equation}
 c_{E^c_3} c_{L_3} = 7.66~10^{-5}
 \qquad \text{and} \qquad
 \frac{c_{L_3}^2}{2 \Lambda_L} = \frac{6.60~10^{-2}}{10^{16}~\text{GeV}}\,.
\end{equation}
Thus, this model can successfully accommodate quark and lepton 
observables for the same value of $\tau$. 
In addition, for this best-fit point we obtain the following predictions in the lepton sector:
normal ordering with\footnote{In the parameterisation of the PMNS matrix we use, 
$U_\mathrm{PMNS} = V P$, where $V$ is a CKM-like matrix containing 
$\theta^\ell_{ij}$ and $\delta_\mathrm{PMNS}$, 
and $P=\mathrm{diag}(1,e^{i\alpha_{21}/2},e^{i\alpha_{31}/2})$ 
containing the Majorana phases $\alpha_{21}$ and $\alpha_{31}$.} 
\begin{align}
 m_1 &\approx 10~\text{meV}\,, \qquad 
  m_{\beta\beta} \approx 6~\text{meV}\,, \qquad
 \sum_{i=1}^{3} m_i \approx 74~\text{meV}\,, \\
 \delta_\mathrm{PMNS}&\approx 0.94\,\pi\,, \qquad
 \alpha_{21} \approx 1.29\,\pi\,, \qquad
 \alpha_{31} \approx 0.14\,\pi\,. 
\end{align} 
We note that the obtained value of $\delta_\mathrm{PMNS}$ is within the experimental $2\sigma$ range~\cite{2503.07752}.

\smallskip

Going beyond the best-fit point(s), we also computed the  confidence regions in the parameter space of the model
using the techniques in~\cite{2409.15823}.
Fig.~\ref{fig:tauregions} shows the $1\sigma$, $2\sigma$ and $3\sigma$ 
regions in the $\tau$ plane obtained by minimising (i)~$\chi^2_q(\tau,\wp_q)$ {(green)}, 
(ii) $\chi^2_\ell(\tau,\wp_\ell)$ {(orange)} (a second CP-conjugated minimum exists)
and (iii) $\chi^2_\mathrm{tot}(\tau,\wp_q,\wp_\ell) = \chi^2_q(\tau,\wp_q) + \chi^2_\ell(\tau,\wp_\ell)$ {(blue)} and marginalising over $\wp_q$ and $\wp_\ell$.
%
The orange solid lines show the $1\sigma$, $2\sigma$ and $3\sigma$ contours for the lepton model with $c^e_{33} = c^\nu_{33} = 441/691$, such that $Y^e_{33} = C^\nu_{33} = E_{12}$. 
A perfect fit to the lepton data only can be obtained for
$ \tau = -0.436 + 0.990$
and
$ l_{13} = 8.22\,, ~ l_{23} = 1.03\,, ~
 e_{13} = 0.306\,, ~ e_{23} = 0.00037\,,
 $
and the overall scales
\begin{equation}
 c_{E^c_3} c_{L_3} = 7.45~10^{-4}
 \qquad \text{and} \qquad
 \frac{c_{L_3}^2}{2 \Lambda_L} = \frac{7.12~10^{-2}}{10^{16}~\text{GeV}}\,.
\end{equation}

\section{Theories with local supersymmetry}
The solution found in the case of rigid supersymmetry can be extended to the supergravity
framework, under few additional hypotheses.
Supergravity actions depend on $K$ and $W$ through the combination:
\be
G=\frac{K}{\bar{M}_{\rm Pl}^2}+\ln\left\vert\frac{W}{\bar{M}_{\rm Pl}^3}\right\vert^2\,.
\ee
To illustrate this more clearly, we adopt the same $K$ and $W$ as in the rigid case, eq.s~(\ref{KandW}) and (\ref{eq:Wgen}).
The rephasing-invariant strong CP angle $\bar\theta$ can again be written in terms of a
rephasing-invariant function 
\beq \bar\theta=\arg \left[e^{-8\pi^2 f_3}  M_3^{C_3}\det Y_u \det Y_d\right]\,, \eeq
that also involves the gluino mass $M_3$, with power given by the QCD adjoint Casimir
$C_3=3$. 
The gluino mass $M_3$ and $W^\dagger$ have the same phase in theories where supersymmetry is broken by modular-invariant dynamics, such as the dilaton.
Under this condition
\beq
\bar\theta=\arg A\qquad\hbox{where now}\qquad A = e^{-8\pi^2 f_3}  W^{-C_3}\det Y_u \det Y_d\,.
\eeq
Under modular transformations the super-potential transforms as
\beq\label{Wmod} W\to (c\tau + d)^{-k_W}W\qquad \hbox{with modular weight}\qquad k_W = \frac{h^2}{\bp^2} . \eeq
This vanishes $k_W\simeq 0$ only in the global supersymmetric limit $h\ll \bp$.
String theory and general arguments motivate integer values for $k_W$. 
Since the superpotential $W$ has negative modular weight $-k_W$,
Yukawa couplings must transform with modular weights 
\beq \label{Ymod} k_{Y^{u}_{ij}}=k_{U^c_i}+k_{Q_j}+k_{H_u}-k_W,\qquad  k_{Y^{d}_{ij}}=k_{D^c_i}+k_{Q_j}+k_{H_d}-k_W.\eeq
The coefficients of mixed modular-gauge anomalies acquire extra terms compared to the rigid limit:
for a generic group factor $G_a$ one has (see e.g.~section 5 of \cite{Feruglio:2023uof})
\beq \label{eq:kfa}
k_{f_a} +\sum_M 2 T_a(M) k_M + k_W \bigg[C_a - \sum_M T_a(M)\bigg] = 0.\eeq
In particular, with the matter field content of the MSSM, the modular-QCD anomaly cancels when
\beq \label{eq:localQCDmodanom}
k_{f_3}+ k_W C_3   + \sum_{i=1}^3  (2k_{Q_i}+ k_{U^c_{i}}+k_{D^c_{i}} - 2k_W)=0. \eeq
From eq.s (\ref{Wmod}), (\ref{Ymod}) and (\ref{eq:localQCDmodanom}) it follows that $A$ is a modular function with weight $k_A=3(k_{H_u}+k_{H_d})$, as in the rigid case. Under the same assumptions made in section~\ref{sec:th},
the angle $\bar\theta$ vanishes.

We can also reproduce the concrete example of section \ref{imod}.
We again set $k_{H_u}=k_{H_d}=0$ and again impose that
the modular weight 
of $\det Y_u\det Y_d$ is $k_Y = 12 m$ with integer $m$.
The case $m=4$ discussed in section \ref{imod} is obtained by modifying the modular weights of quarks e.g.\ into
\be
\label{eq:Qweights2}
k_{U^c_i}=k_{D^c_i}=k_{Q_i}=(2,4,6) + k_W/2.
\ee
In this case
\beq
e^{-8\pi^2 f_3}  W^{-C_3}\propto \Delta(\tau)^{-4}\,,
\eeq
where the proportionality can involve a dependence on $S$.
This behaviour is compatible with a singularity at $\tau=i\infty$ exhibited by both $f_3$, as seen in the case of global supersymmetry, and by the super-potential $W$. The latter singularity is actually implied by being $W$ a modular function
of negative weight, eq.~(\ref{Wmod}). For example the desired properties of the super-potential are reproduced if $\langle W\rangle=c(S)/\eta^{2 k_W}(\tau)$. For consistency, under modular transformations the whole super-potential $W$ should transform as $\eta^{-2 k_W}(\tau)$, which may require the laws of matter fields to involve non-trivial multipliers.

\section{Implications on unification of gauge couplings}\label{sec:GUT}
We now examine the implications of our assumptions --- motivated by the strong CP problem --- for the unification of gauge coupling constants, one of the few available indirect probes of physics near the Planck scale.
The two issues are related because the supersymmetric gauge kinetic functions of eq.~(\ref{susylag}) contain in their real and imaginary parts
the gauge coupling $g_a$ and the $\theta_a$ angle of each gauge group
\be
f_a =\frac{1}{g_a^2}-i\frac{\theta_a}{8\pi^2}.
\ee
The weak angles $\theta_{1,2}$ can be rotated away.
The gauge couplings $g_a$ are known as `holomorphic' because written in a field basis
where the holomorphicity properties of supersymmetric theories are explicit,
at the price of having non-canonically normalised vector and matter fields.
In theories with global supersymmetry, the transformations needed to reach the basis where all fields are canonical are
\begin{align}
\label{gocan}
M\to Z_M^{-1/2} M , \qquad
V_a\to g_a^c V_a\,,
\end{align}
where $g_a^c$ denote the `physical' gauge couplings.
In theories with local supersymmetry, an additional transformation is needed. Indeed the matter-coupled supergravity Lagrangian is usually formulated in the Brans-Dicke form, where the graviton kinetic term arising from the
scalar curvature ${\cal R}$ has a non-canonical coupling to matter scalars:
\be
{\cal E}^{-1}{\cal L}_{\rm BD}=-\frac{1}{2} e^{-K/3\bar{M}_{\rm Pl}^2}{\cal R}+....
\ee
Here ${\cal E}$ is the vielbein determinant, ${\cal R}$ the scalar curvature.
A super-Weyl transformation is required to transform it into the Einstein form
\be
{\cal E}^{-1}{\cal L}_{E}=-\frac{1}{2}{\cal R}+ \cdots.
\ee
After rescaling the graviton, kinetic terms of fermions also need to get
rescaled by exponentials of $K/\bar{M}_{\rm Pl}^2$~\cite{Kaplunovsky:1995jw},
such that  bosons and fermions belonging to the same supermultiplet are normalized in the same way.
In general, these field redefinitions --- involving a rescaling of all fermions --- are anomalous and
the associated anomaly~\cite{Konishi:1985tu,Arkani-Hamed:1997qui,Kaplunovsky:1995jw} produces a one-loop shift of the real part of the gauge kinetic functions. 
As a result the physical gauge coupling constants $g_a^c$ get related to the holomorphic couplings by the following
implicit relation
\begin{align}
\label{implicit}
\frac{1}{g_a^{c2}}=&~\frac{1}{g_a^2}-\frac{1}{8\pi^2} \sum_M T_a(M) \ln Z_M- \frac{1}{8\pi^2} C_a \ln  g^{c2}_a+\frac{1}{16\pi^2}\left[-C_a+\sum_M T_a(M)\right]\frac{K}{\bar{M}_{\rm Pl}^2}\,,
\end{align}
where $C_{a}$ are the quadratic Casimir of SU$(a)$ $(a=2,3)$ and $C_1=0$.
The physical gauge couplings $g^c_a$ are modular invariant, unlike the holomorphic couplings $g_a$.\footnote{
This intuitively expected result follows from 
$f_a\to f_a-({k_{f_a}}/{8\pi^2}) \ln (c\tau+d)$, 
$K\to K+k_W \bar{M}_{\rm Pl}^2 \ln |c\tau+d|^2$,
$\ln Z_M\to \ln Z_M+ k_M \ln |c\tau+d|^2$ and 
$y\to y |c\tau+d|^{-2}$.}
As already discussed, modular invariance implies that $g_a$ must receive extra loop corrections that,
combined with the correction arising from the anomaly, form a modular-invariant quantity.
In string theory these corrections are provided by the heavy modes~\cite{Kaplunovsky:1995jw}.

An additional property of the holomorphic coupling $g_a$ is that it only runs at 1-loop~\cite{Shifman:1991dz}:
\begin{align}
\label{oneloop}
\frac{1}{g_a^2(\mu)}=\frac{1}{g_a^2(\Lambda)}+\frac{b_a}{16\pi^2} \ln \frac{{\Lambda}^2}{\mu^2}\qquad\hbox{with $\beta$ function}\qquad
b_a=-3 C_a+\sum_M T_a(M)\,.
\end{align}
Eq.~(\ref{implicit}) holds at a generic scale $\mu$ and 
by combining it with eq.~(\ref{oneloop}) an all-loop beta function for the physical couplings $g^c_a(\mu)$ is obtained~\cite{Novikov:1983uc,Novikov:1985rd}. 

To illustrate the modular-invariant result of eq.~(\ref{implicit}) with an explicit example incorporating the RGE flow, 
we work at one-loop order and we adopt the choice of eq.~(\ref{KandW}) together with
\begin{align}
K_S=- {\bp}^2\ln(S+\bar S)\,, \qquad f_a=\kappa_a S-\frac{k_{f_a}}{8\pi^2}\ln \eta^2(\tau)\,. \label{eq:KS}
\end{align}
In this case we get\footnote{That is, we 
take $g_a^{-2}(\Lambda)= k_a \Re S-k_{f_a}/8\pi^2 \ln |\eta|^2$, $Z_M(\mu)\approx Z_M(\Lambda)=y^{-k_M}$, $K(\mu)\approx K(\Lambda)=-\bar{M}_{\rm Pl}^2\ln(S+\bar S)-k_W \bar{M}_{\rm Pl}^2 \ln y$
and $1/g_a^{c2}\approx \kappa_a \Re S$ in the right-hand-side of eq.~(\ref{implicit}). }
\begin{align}
\label{solthre}
\frac{16\pi^2}{g_a^{c2}(\mu)}=16 \pi^2 \kappa_a \Re S+2 C_a \ln \frac{\kappa_a}{2}
+b_a \ln \frac{{\Lambda}^2}{2 \Re S\, \mu^2}
+\Delta_a(\tau,\bar\tau)\,,
\end{align}
where  the threshold corrections
\begin{align}
\Delta_a(\tau,\bar\tau)\equiv -k_{f_a}\ln y|\eta(\tau)|^4
\end{align}
are entirely determined by the weights of the matter multiplets and by the transformation properties of $f_a(S,\tau)$. 
Threshold corrections $\Delta_a$ are positive for $k_{f_a}<0$ and  get large for $\tau\to i\infty$.
To interpret this result and go beyond QFT considerations, we recall that
corrections of this type arise in string theory from the contribution of the stringy heavy modes~\cite{hep-th/9109053,Derendinger:1991hq,Ferrara:1990ei,Dixon:1990pc,Kaplunovsky:1994fg,Kaplunovsky:1995jw,hep-th/9602045}. 
In this string context
\begin{itemize}
\item $\tau\to i \infty$ is a de-compactification limit, where extra dimensions get large.
Summing the effects of all ${\rm SL}(2,\mathbb{Z})^2$ toroidal moduli reproduces, in the decompactification limits, 
the running of gauge couplings in extra dimensions. 
In our string-like context the extra-dimensional $\beta$ function coefficients are determined by modular weights
of the MSSM fields, while in~\cite{hep-ph/9806292} they were determined summing over Kaluza Klein modes.

\begin{figure}[t]
$$\includegraphics[width=0.65\textwidth]{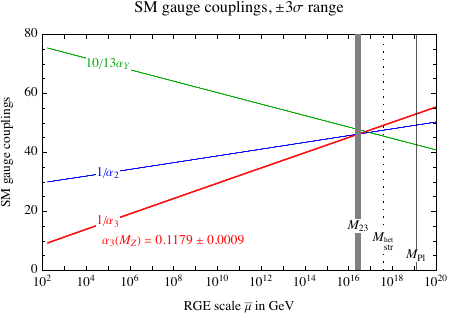}$$
\begin{center}
\caption{\em\label{fig:SMgaugeStringUnification}
Three-loop running of the SM gauge couplings.
The gray band shows the energy at which $\alpha_2(\mu)=\alpha_3(\mu)$, a factor of 10 below than the unification scale predicted at tree level by heterotic strings.
}
\end{center}
\end{figure}

\item
The QFT cut-off $\Lambda$ becomes the string unification scale.
It is predicted to be $ g_a \bp \sim \hbox{few}~10^{17}\,{\rm GeV}$ in heterotic strings where gauge and gravitational interactions
arise from the same open-string sector, but other constructions avoid this prediction~\cite{hep-th/9602045}.

\item Non-abelian groups $G_a$ have integer Kac-Moody $\kappa_a$, while $\kappa_1$ for hypercharge
can be a rational number.
\end{itemize}
As no full string model is available, we try proceeding by mixing string and QFT considerations.
In our minimal model, quarks and leptons of each family have equal modular weights and form full SU(5) multiplets,
contributing to $k_{f_a}$ in eq.\eq{kfa} as $-48 \tilde\kappa_a $
where $\tilde{\kappa}_{2,3}=\kappa_{2,3}$ and $\tilde{\kappa}_1 = 3\kappa_1/5$ equals 1 in GUT normalization.
The Higgses $H_{u,d}$ do not contribute to $k_{f_a}$ as we assumed modular weights  $k_{H_u}+k_{H_d}=0$.
Including the supergravity gauge terms gives 
\beq -k_{f_a} = 48\tilde \kappa_a + k_W (C_a - 6\tilde\kappa_a).\eeq
The terms proportional to $\kappa_a$ give threshold corrections  that can be reabsorbed in a redefinition of the unified gauge coupling.
The term proportional to $C_a$ affects unification.

A simple possibility is having SU(5)-like Kac-Moody values $\kappa_3=\kappa_2=1$ and $\kappa_1 = 5/3$.
As well known, the MSSM gauge couplings unify around $2~10^{16}\,{\rm GeV}$ provided that 
supersymmetry exists around a low scale not much above the weak scale.
A more precise discussion is not possible because all masses of supersymmetric particles are unknown.

The opposite limit of high-scale supersymmetry provides another minimal possibility.
Fig.\fig{SMgaugeStringUnification} shows the running of gauge couplings in the SM.
Given that all its parameters have been measured, the running can be computed precisely at 3 loop accuracy.
The main uncertainty is on the strong coupling.
The SU(2) and SU(3) gauge couplings cross at a scale $M_{23}\approx 3~10^{16}\,{\rm GeV}$.
The threshold corrections proportional to $C_a$ increase $M_{23}$, but only by a factor of 2.
To unify at the same scale, the hyper-charge Kac-Moody level needs to acquire a value such as $\kappa_1\sim 13/10$~\cite{hep-th/9602045}.
Extra corrections are provided by supersymmetric particles, neglected in the figure.

\section{Conclusions}
We have extended a previously proposed modular solution to the strong CP problem~\cite{Feruglio:2023uof} 
to effective field theories inspired by string compactifications, 
which generically involve quarks with positive modular weights and non-trivial gauge kinetic functions. 
Under the assumptions that CP is only broken by the modulus $\tau$ of a ${\rm SL}(2,\mathbb{Z})$ modular symmetry,
that the relevant modular functions are holomorphic in the fundamental domain (justified by supersymmetry)
and that singularities appear only in the limit $\tau\to i \infty$
(motivated as de-compactification limit  of some string compactifications, and by more general string `distance' conjectures),
we demonstrate that $\tau$ can generate a large CKM phase without generating a QCD theta angle $\bar\theta$.
This is possible because $\bar\theta$ is the phase of a chiral superfield --- $e^{-8\pi^2 f_3} \det Y_u \det Y_d$ in case of rigid supersymmetry --- where each term transforms with a modular weight. 
The first gauge factor is singular, and we assumed
the QCD kinetic function predicted by a class of string compactifications with a full ${\rm SL}(2,\mathbb{Z})$ modular invariance.
To achieve $\bar\theta=0$ we assumed a specific structure of quark Yukawa matrices, such that their determinant is proportional to a
power of the modular discriminant, the lowest-weight modular form that vanishes for $\tau\to i \infty$ only.
Eq.~(\ref{deus}) exemplifies the needed Yukawa matrices, assuming modular weights $(2,4,6)$ for the three quark generations.
The assumed pattern of Yukawa couplings does not appear to be motivated by semi-realistic string computations.
Instead, it  accommodates all observed quark masses, mixing angles and CKM phase for the value of $\tau$ fitted in fig.\fig{tauregions}.
Assuming that leptons have the same $(2,4,6)$ modular weights, and allowing for more general Yukawa matrices, allows us
to reproduce also lepton masses and mixings for the same $\tau$. 

Following the string motivation, we also included the dilaton $S$. 
The imaginary part of $S$ behaves as an axion. 
The theory has a continuous shift symmetry, $\Im S\to \Im S+\mathrm{constant}$, broken by the coupling to the gauge vector bosons, see eq. \eqref{eq:KS}.
In our notation, $\Im S$ includes the contribution of the QCD vacuum.
In a full-fledged theory, the VEVs of $S$ and $\tau$ should be determined dynamically.
In the explicit model we discuss in section \ref{imod}, $\langle\tau\rangle$ is obtained by fitting fermion
masses and mixing angles, while $\langle S\rangle$ is real by assumption.
In a CP-invariant theory, CP-preserving points are extrema of the energy density, and were long
conjectured to be local minima~\cite{Ferrara:1990ei,Cvetic:1991qm}. Only recently, minima with CP-violating
values of $\langle\tau\rangle$ have been found~\cite{Novichkov:2022wvg,Leedom:2022zdm}.
Concerning the VEV of $S$, two options are open. 
On the one hand, its imaginary part can be stabilised to zero by QCD dynamics, thus reproducing the QCD axion solution. 
On the other hand, the VEV of $S$ can be real as the result of Planck-scale dynamics. An example of this second possibility has been recently discussed in \cite{2405.18813}.
See also~\cite{hep-th/0510053} for a different mechanism based on accumulation of vacua around $\bar\theta=0$.

When supersymmetry breaking effects are turned on, $\bar\theta$ can deviate from zero. The size of the corrections can be kept under control, depending on the scale and the mechanism of supersymmetry breaking~\cite{Hiller:2001qg,Hiller:2002um,Feruglio:2023uof}.

Finally, we discussed how these structures extend to local supersymmetry and the mild implications for string gauge coupling unification.
We have not here considered string compactifications invariant under a modular sub-group,
allowing for lower modular weights $k \sim 1$,
but perhaps weakening the modular control on the QCD angle~\cite{hep-th/9303017}.


\small

\paragraph{Note added}
The authors of \cite{2505.08358} argue that:
i) in the QCD action $\theta_{\rm QCD}$ is the coefficient of a non-dynamical total derivative, implying that it selects a state of the theory without affecting its dynamics.
ii) Solutions to the strong CP problem based on P or CP symmetries are not valid. 
This would imply that the strong CP problem exists even in a CP-conserving theory with vanishing  CKM phase.
iii) The axion is a valid solution to the strong CP problem as $a G \tilde{G}$ with a dynamical scalar $a(x)$
no longer is a total derivative.

Modular invariant theories too add a dynamical scalar $\tau(x)$ such that $f_3(\tau) G\tilde{G}$ is not a total derivative.
However, we assumed special modular charges such that $\bar\theta$ does not depend on $\tau$, that instead generates the CKM phase.
So, this is equivalent to having reduced the SM strong CP problem to a SM-like theory with vanishing phase of the determinant of the quark mass matrix. 
If this solution is incomplete, as argued in~\cite{2505.08358}, it can be easily completed:
in the class of theories we considered $\theta_{\rm QCD}$ anyhow remains controlled by dynamical super-fields, such as the dilaton $S$.
One can easily imagine CP-conserving Planck-scale dynamics such that the potential has a minimum at some real $S$.
More in general, $\theta_{\rm QCD}$ and the strong gauge coupling get unified in an $S$ super-multiplet,
making clear that, unlike in QCD, $\theta_{\rm QCD}$ no longer is the coefficient of a total derivative that does not affect the Hamiltonian.


\footnotesize

\paragraph{Acknowledgements.} 
We thank T. Gherghetta, A.\ Kobakhidze, P.\ Nilles, S.\ Rajendran, M.\ Reig and M.\ Zantedeschi for discussions.
A.T.\ is funded by the European Union, Next Generation EU, 
National Recovery and Resilience Plan
(mission~4, component~2) 
under the project \textit{MODIPAC: Modular Invariance in Particle Physics and Cosmology} (CUP~C93C24004940006).
The work of A.M. was partially supported by the research grant number 2022E2J4RK ``PANTHEON: Perspectives in Astroparticle and Neutrino THEory with Old and New messengers'' under the program PRIN 2022 funded by the Italian Ministero dell'Universit\`a e della Ricerca (MUR) and by the European Union -- Next Generation EU, as well as by the Theoretical Astroparticle
Physics (TAsP) initiative of the Istituto Nazionale di Fisica Nucleare (INFN). 
F.F. is supported by INFN.

\bibliographystyle{JHEP2}

\footnotesize
\begin{thebibliography}{nnn}\bibitem{Feruglio:2023uof}
\article[2305.08908]{F. Feruglio, A. Strumia, A. Titov}{JHEP}{07}{027}{2023}
{\href{https://doi.org/10.1007/JHEP07(2023)027}{Modular invariance and the QCD angle}}.


\bibitem{2404.08032}
\article[2404.08032]{J.T. Penedo, S.T. Petcov}{JHEP}{10}{172}{2024}
{\href{https://doi.org/10.1007/JHEP10(2024)172}{Finite modular symmetries and the strong CP problem}}.


\bibitem{2406.01689}
\article[2406.01689]{F. Feruglio, M. Parriciatu, A. Strumia, A. Titov}{JHEP}{08}{214}{2024}
{\href{https://doi.org/10.1007/JHEP08(2024)214}{Solving the strong CP problem without axions}}.


\bibitem{Kaplunovsky:1995jw}
\article[hep-th/9502077]{V. Kaplunovsky, J. Louis}{Nucl.Phys.B}{444}{191}{1995}
{\href{https://doi.org/10.1016/0550-3213(95)00172-O}{On Gauge couplings in string theory}}.


\bibitem{1307.0710}
\article[1307.0710]{S. Antusch, M. Holthausen, M.A. Schmidt, M. Spinrath}{Nucl.Phys.B}{877}{752}{2013}
{\href{https://doi.org/10.1016/j.nuclphysb.2013.10.028}{Solving the Strong CP Problem with Discrete Symmetries and the Right Unitarity Triangle}}.


\bibitem{1412.3805}
\article[1412.3805]{L. Vecchi}{JHEP}{04}{149}{2017}
{\href{https://doi.org/10.1007/JHEP04(2017)149}{Spontaneous CP violation and the strong CP problem}}.


\bibitem{1506.05433}
\article[1506.05433]{M. Dine, P. Draper}{JHEP}{08}{132}{2015}
{\href{https://doi.org/10.1007/JHEP08(2015)132}{Challenges for the Nelson-Barr Mechanism}}.


\bibitem{2105.09122}
\article[2105.09122]{A. Valenti, L. Vecchi}{JHEP}{07}{203}{2021}
{\href{https://doi.org/10.1007/JHEP07(2021)203}{The CKM phase and $ \overline{\theta} $ in Nelson-Barr models}}.


\bibitem{2106.09108}
\article[2106.09108]{A. Valenti, L. Vecchi}{JHEP}{07}{152}{2021}
{\href{https://doi.org/10.1007/JHEP07(2021)152}{Super-soft CP violation}}.


\bibitem{2406.01260}
\article[2406.01260]{S. Nakagawa, Y. Nakai, Y. Wang}{JHEP}{09}{105}{2024}
{\href{https://doi.org/10.1007/JHEP09(2024)105}{Spontaneous CP violation in supersymmetric QCD}}.


\bibitem{2407.14585}
\heparticle[2407.14585]{L. Hall, C.A. Manzari, B. Noether}{Strong CP and Flavor in Multi-Higgs Theories}.


\bibitem{2407.18161}
\article[2407.18161]{R. Ferro-Hernandez, S. Morisi, E. Peinado}{Phys.Rev.D}{111}{073009}{2025}
{\href{https://doi.org/10.1103/PhysRevD.111.073009}{Axionless strong CP problem solution: The spontaneous CP-violation case}}.


\bibitem{2408.12146}
\article[2408.12146]{Q. Liang, R. Okabe, T.T. Yanagida}{Phys.Lett.B}{859}{139123}{2024}
{\href{https://doi.org/10.1016/j.physletb.2024.139123}{Three-zero texture of quark-mass matrices as a solution to the strong CP problem}}.


\bibitem{2505.05142}
\heparticle[2505.05142]{Q. Liang, T.T. Yanagida}{Non-invertible symmetry as an axion-less solution to the strong CP problem}.


\bibitem{2405.18813}
\article[2405.18813]{T. Higaki, T. Kobayashi, K. Nasu, H. Otsuka}{JHEP}{09}{024}{2024}
{\href{https://doi.org/10.1007/JHEP09(2024)024}{Spontaneous CP violation and partially broken modular flavor symmetries}}.


\bibitem{1905.11970}
\article[1905.11970]{P.P. Novichkov, J.T. Penedo, S.T. Petcov, A.V. Titov}{JHEP}{07}{165}{2019}
{\href{https://doi.org/10.1007/JHEP07(2019)165}{Generalised CP Symmetry in Modular-Invariant Models of Flavour}}.


\bibitem{Ooguri:2006in}
\article[hep-th/0605264]{H. Ooguri, C. Vafa}{Nucl.Phys.B}{766}{21}{2007}
{\href{https://doi.org/10.1016/j.nuclphysb.2006.10.033}{On the Geometry of the String Landscape and the Swampland}}.


\bibitem{Gonzalo:2018guu}
\article[1812.06520]{E. Gonzalo, L.E. Ib{\' a}{\~ n}ez, {\' A}.M. Uranga}{JHEP}{05}{105}{2019}
{\href{https://doi.org/10.1007/JHEP05(2019)105}{Modular symmetries and the swampland conjectures}}.


\bibitem{Ferrara:1991uz}
\article{S. Ferrara, C. Kounnas, D. Lust, F. Zwirner}{Nucl.Phys.B}{365}{431}{1991}
{\href{https://doi.org/10.1016/S0550-3213(05)80028-8}{Duality invariant partition functions and automorphic superpotentials for (2,2) string compactifications}}.


\bibitem{Antusch:2013jca}
\article[1306.6879]{S. Antusch, V. Maurer}{JHEP}{11}{115}{2013}
{\href{https://doi.org/10.1007/JHEP11(2013)115}{Running quark and lepton parameters at various scales}}.


\bibitem{2503.07752}
\heparticle[2503.07752]{F. Capozzi, W. Giar{\` e}, E. Lisi, A. Marrone, A. Melchiorri, A. Palazzo}{Neutrino masses and mixing: Entering the era of subpercent precision}.


\bibitem{2409.15823}
\article[2409.15823]{G.-J. Ding, E. Lisi, A. Marrone, S.T. Petcov}{Phys.Rev.D}{111}{075024}{2025}
{\href{https://doi.org/10.1103/PhysRevD.111.075024}{Interplay and correlations between quark and lepton observables in modular symmetry models}}.


\bibitem{Konishi:1985tu}
\article{K. Konishi, K. Shizuya}{Nuovo Cim.A}{90}{111}{1985}
{\href{https://doi.org/10.1007/BF02724227}{Functional Integral Approach to Chiral Anomalies in Supersymmetric Gauge Theories}}.


\bibitem{Arkani-Hamed:1997qui}
\article[hep-th/9707133]{N. Arkani-Hamed, H. Murayama}{JHEP}{06}{030}{2000}
{\href{https://doi.org/10.1088/1126-6708/2000/06/030}{Holomorphy, rescaling anomalies and exact beta functions in supersymmetric gauge theories}}.


\bibitem{Shifman:1991dz}
\article{M.A. Shifman, A.I. Vainshtein}{Nucl.Phys.B}{359}{571}{1991}
{\href{https://doi.org/10.1016/0550-3213(91)90072-6}{On holomorphic dependence and infrared effects in supersymmetric gauge theories}}.


\bibitem{Novikov:1983uc}
\article{V.A. Novikov, M.A. Shifman, A.I. Vainshtein, V.I. Zakharov}{Nucl.Phys.B}{229}{381}{1983}
{\href{https://doi.org/10.1016/0550-3213(83)90338-3}{Exact Gell-Mann-Low Function of Supersymmetric Yang-Mills Theories from Instanton Calculus}}.


\bibitem{Novikov:1985rd}
\article{V.A. Novikov, M.A. Shifman, A.I. Vainshtein, V.I. Zakharov}{Phys.Lett.B}{166}{329}{1986}
{\href{https://doi.org/10.1016/0370-2693(86)90810-5}{The beta function in supersymmetric gauge theories. Instantons versus traditional approach}}.


\bibitem{hep-th/9109053}
\article[hep-th/9109053]{L.E. Ibanez, D. Lust, G.G. Ross}{Phys.Lett.B}{272}{251}{1991}
{\href{https://doi.org/10.1016/0370-2693(91)91828-J}{Gauge coupling running in minimal SU(3) $\otimes$ SU(2) $\otimes$ U(1) superstring unification}}.


\bibitem{Derendinger:1991hq}
\article{J.P. Derendinger, S. Ferrara, C. Kounnas, F. Zwirner}{Nucl.Phys.B}{372}{145}{1992}
{\href{https://doi.org/10.1016/0550-3213(92)90315-3}{On loop corrections to string effective field theories: Field dependent gauge couplings and sigma model anomalies}}.


\bibitem{Ferrara:1990ei}
\article{S. Ferrara, N. Magnoli, T.R. Taylor, G. Veneziano}{Phys.Lett.B}{245}{409}{1990}
{\href{https://doi.org/10.1016/0370-2693(90)90666-T}{Duality and supersymmetry breaking in string theory}}.


\bibitem{Dixon:1990pc}
\article{L.J. Dixon, V. Kaplunovsky, J. Louis}{Nucl.Phys.B}{355}{649}{1991}
{\href{https://doi.org/10.1016/0550-3213(91)90490-O}{Moduli dependence of string loop corrections to gauge coupling constants}}.


\bibitem{Kaplunovsky:1994fg}
\article[hep-th/9402005]{V. Kaplunovsky, J. Louis}{Nucl.Phys.B}{422}{57}{1994}
{\href{https://doi.org/10.1016/0550-3213(94)00150-2}{Field dependent gauge couplings in locally supersymmetric effective quantum field theories}}.


\bibitem{hep-th/9602045}
\article[hep-th/9602045]{K.R. Dienes}{Phys.Rept.}{287}{447}{1997}
{\href{https://doi.org/10.1016/S0370-1573(97)00009-4}{String theory and the path to unification: a review of recent developments}}.


\bibitem{hep-ph/9806292}
\article[hep-ph/9806292]{K.R. Dienes, E. Dudas, T. Gherghetta}{Nucl.Phys.B}{537}{47}{1999}
{\href{https://doi.org/10.1016/S0550-3213(98)00669-5}{Grand unification at intermediate mass scales through extra dimensions}}.


\bibitem{Cvetic:1991qm}
\article{M. Cvetic, A. Font, L.E. Ibanez, D. Lust, F. Quevedo}{Nucl.Phys.B}{361}{194}{1991}
{\href{https://doi.org/10.1016/0550-3213(91)90622-5}{Target space duality, supersymmetry breaking and the stability of classical string vacua}}.


\bibitem{Novichkov:2022wvg}
\article[2201.02020]{P.P. Novichkov, J.T. Penedo, S.T. Petcov}{JHEP}{03}{149}{2022}
{\href{https://doi.org/10.1007/JHEP03(2022)149}{Modular flavour symmetries and modulus stabilisation}}.


\bibitem{Leedom:2022zdm}
\article[2212.03876]{J.M. Leedom, N. Righi, A. Westphal}{JHEP}{02}{209}{2023}
{\href{https://doi.org/10.1007/JHEP02(2023)209}{Heterotic de Sitter beyond modular symmetry}}.


\bibitem{hep-th/0510053}
\article[hep-th/0510053]{G. Dvali}{Phys.Rev.D}{74}{025019}{2006}
{\href{https://doi.org/10.1103/PhysRevD.74.025019}{A Vacuum accumulation solution to the strong CP problem}}.


\bibitem{Hiller:2001qg}
\article[hep-ph/0105254]{G. Hiller, M. Schmaltz}{Phys.Lett.B}{514}{263}{2001}
{\href{https://doi.org/10.1016/S0370-2693(01)00814-0}{Solving the Strong CP Problem with Supersymmetry}}.


\bibitem{Hiller:2002um}
\article[hep-ph/0201251]{G. Hiller, M. Schmaltz}{Phys.Rev.D}{65}{096009}{2002}
{\href{https://doi.org/10.1103/PhysRevD.65.096009}{Strong Weak CP Hierarchy from Nonrenormalization Theorems}}.


\bibitem{hep-th/9303017}
\article[hep-th/9303017]{P. Mayr, S. Stieberger}{Nucl.Phys.B}{407}{725}{1993}
{\href{https://doi.org/10.1016/0550-3213(93)90096-8}{Threshold corrections to gauge couplings in orbifold compactifications}}.


\bibitem{2505.08358}
\heparticle[2505.08358]{D.E. Kaplan, T. Melia, S. Rajendran}{What can solve the Strong CP problem?}.


\end{thebibliography}

\end{document}